\documentclass[Physsubmission, Phys]{SciPost}

\hypersetup{
    colorlinks,
    linkcolor={red!50!black},
    citecolor={blue!50!black},
    urlcolor={blue!80!black}
}

\usepackage[bitstream-charter]{mathdesign}
\DeclareSymbolFont{usualmathcal}{OMS}{cmsy}{m}{n}
\DeclareSymbolFontAlphabet{\mathcal}{usualmathcal}

\newcommand{\bl}{\mbox{\boldmath $l$}}
\newcommand{\bp}{\mbox{\boldmath $p$}}
\newcommand{\bq}{\mbox{\boldmath $q$}}

\newcommand{\bk}{\mbox{\boldmath $k$}}

\newcommand{\half}{{1\over 2}}

\newcommand{\bQQ}{\mbox{\boldmath $Q$}}

\begin{document}

\begin{center}{\Large \textbf{
Central exclusive production of $\eta_c$ and $\chi_{c0}$ in the \\ light-front k$_{\perp}$-factorization approach \\
}}\end{center}

\begin{center}
Izabela Babiarz\textsuperscript{1$\star$},
Roman Pasechnik \textsuperscript{2} 
Wolfgang Sch\"afer\textsuperscript{1}, and
Antoni Szczurek \textsuperscript{1}
\end{center}

\begin{center}
{\bf 1} Institute of Nuclear Physics Polish Academy of Sciences, Krak\'ow, Poland
\\
{\bf 2} Department of Astronomy and Theoretical Physics, Lund University, SE-223 62 Lund, Sweden
\\
* Izabela.Babiarz@ifj.edu.pl
\end{center}

\begin{center}
\today
\end{center}


\definecolor{palegray}{gray}{0.95}
\begin{center}
\colorbox{palegray}{
  \begin{tabular}{rr}
  \begin{minipage}{0.1\textwidth}
    \includegraphics[width=22mm]{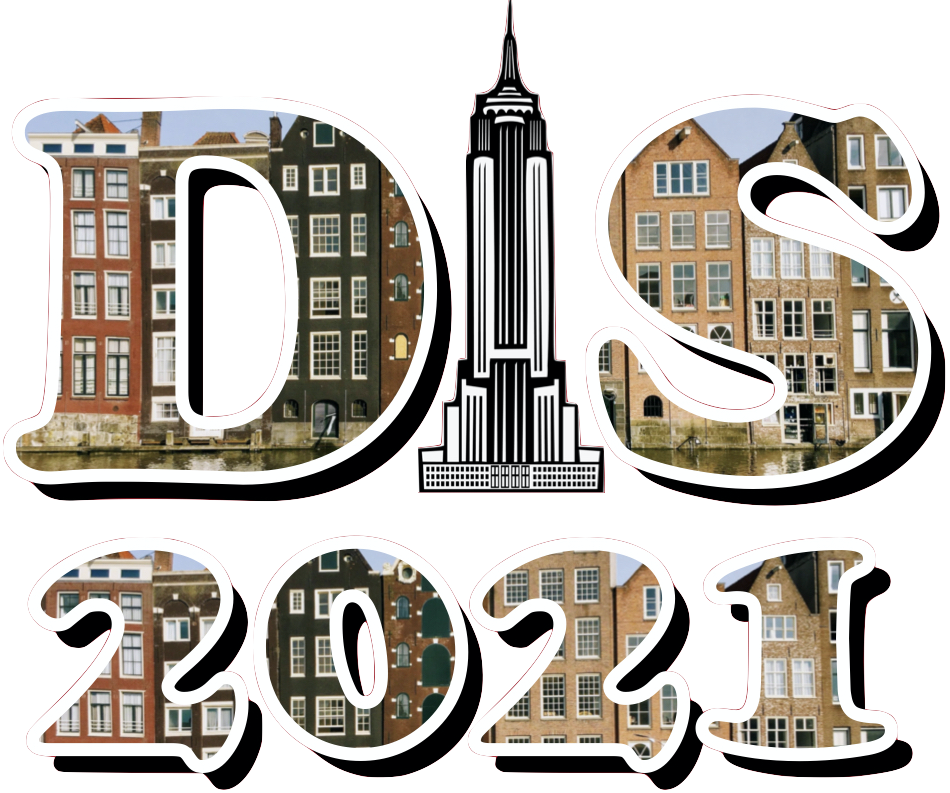}
  \end{minipage}
  &
  \begin{minipage}{0.75\textwidth}
    \begin{center}
    {\it Proceedings for the XXVIII International Workshop\\ on Deep-Inelastic Scattering and
Related Subjects,}\\
    {\it Stony Brook University, New York, USA, 12-16 April 2021} \\
    \doi{10.21468/SciPostPhysProc.?}\\
    \end{center}
  \end{minipage}
\end{tabular}
}
\end{center}

\section*{Abstract}
{\bf
We study the exclusive production of $J^{PC}=0^{++}, 0^{--}$ charmonium states in proton-proton collisions at the LHC energies. 
The $pp \to pp\eta_c$ reaction is discussed for the first time. 
We observe a substantial contribution from the nonperturbative domain of gluon virtualities, especially for $\eta_c$ production. To model the nonperturbative region better, we utilize models of the unintegrated gluon distribution based on parametrizations of the color dipole cross-section.
}


\section{Introduction}
\label{sec:intro}
Central exclusive diffractive processes are distinguished by their very unusual final states. 
The diffractively excited system, e.g. a meson or a few-particle state is produced in the central rapidity region and is fully measured. There are no other tracks 
in the detectors, except perhaps the tagged final state protons. 
Beyond the fully exclusive or ``elastic'' diffraction where the incoming protons remain intact, in absence of proton tagging also 
``inelastic diffraction'' must be accounted for, where small mass hadronic systems, disappear into the beam pipe.
Here we give a brief summary of our recent work \cite{Babiarz:2020jhy} where we have considered two such reactions, $ pp\:\rightarrow p\:\chi_{c0}\:\:p$ 
and $pp\:\rightarrow p\:\eta_{c}\:\:p$.
The final state mesons being composed of heavy (charm) quarks, these processes appear to be well suited to be analysed in the framework of 
the perturbative QCD (pQCD) based on the ``Durham model'' formulated by Khoze, Martin and Ryskin (see Ref.~\cite{Harland-Lang:2014lxa} and references 
therein). 
Building upon the Durham formulation, the theory of the central exclusive production (CEP) of single $\chi_{cJ}$, $J=0,1,2$ mesons,
with a correct account for the spin of the mesons and precise kinematics of the process has been worked out by Pasechnik, Szczurek and Teryaev (PST) in a series 
of papers \cite{Pasechnik:2007hm,Pasechnik:2009bq,Pasechnik:2009qc}. Here we review our recent work, where we revisited and extended this analysis to account 
for additional effects and sources for theoretical uncertainties (such as the shapes 
of the charmonia wave functions).
Also, for the first time, we studied the pseudoscalar $\eta_c$ final state.

\section{Formalism and Results}

\subsection{pQCD description of central exclusive diffraction}

\begin{figure}[h]
\centering
\includegraphics[width=0.5\textwidth]{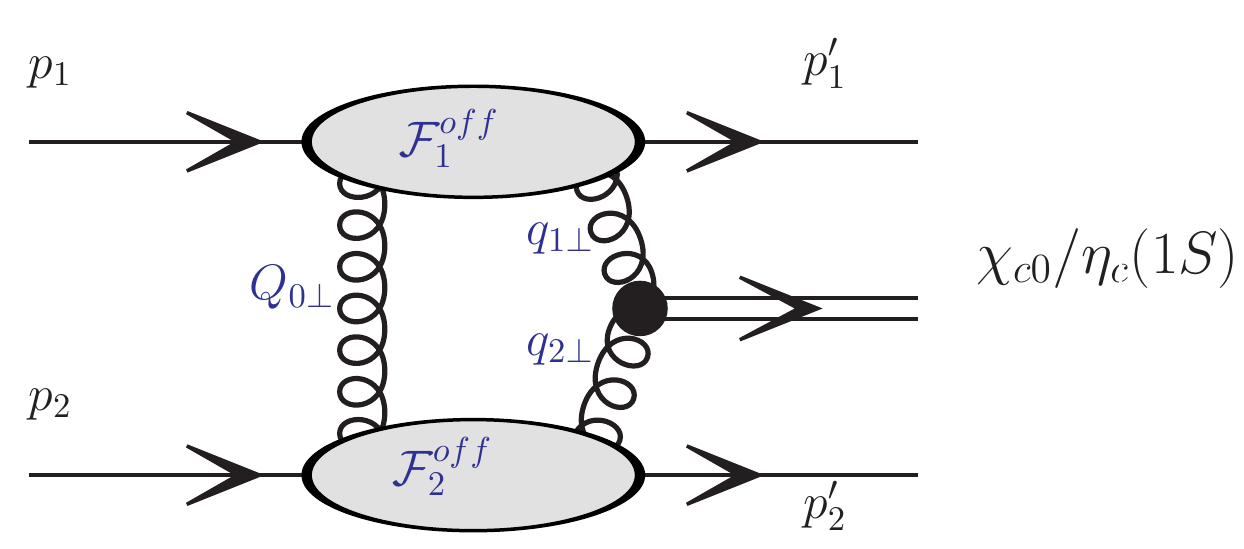}
\caption{A Feynman diagram for the CEP process.}
\label{fig:diagram}
\end{figure}

\begin{figure}
    \centering
    a)\includegraphics[width = 0.4\textwidth]{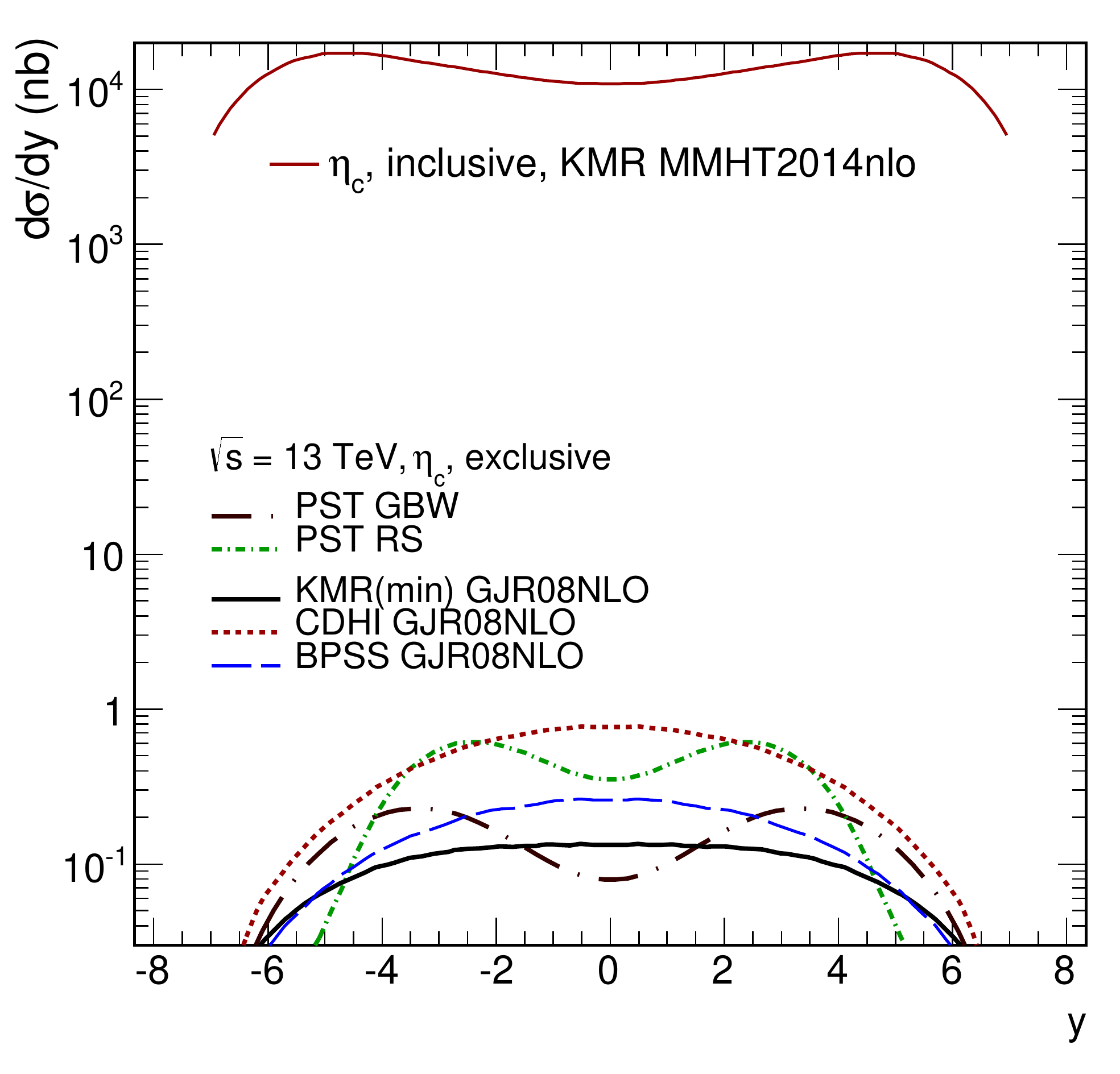}
    b)\includegraphics[width = 0.4\textwidth]{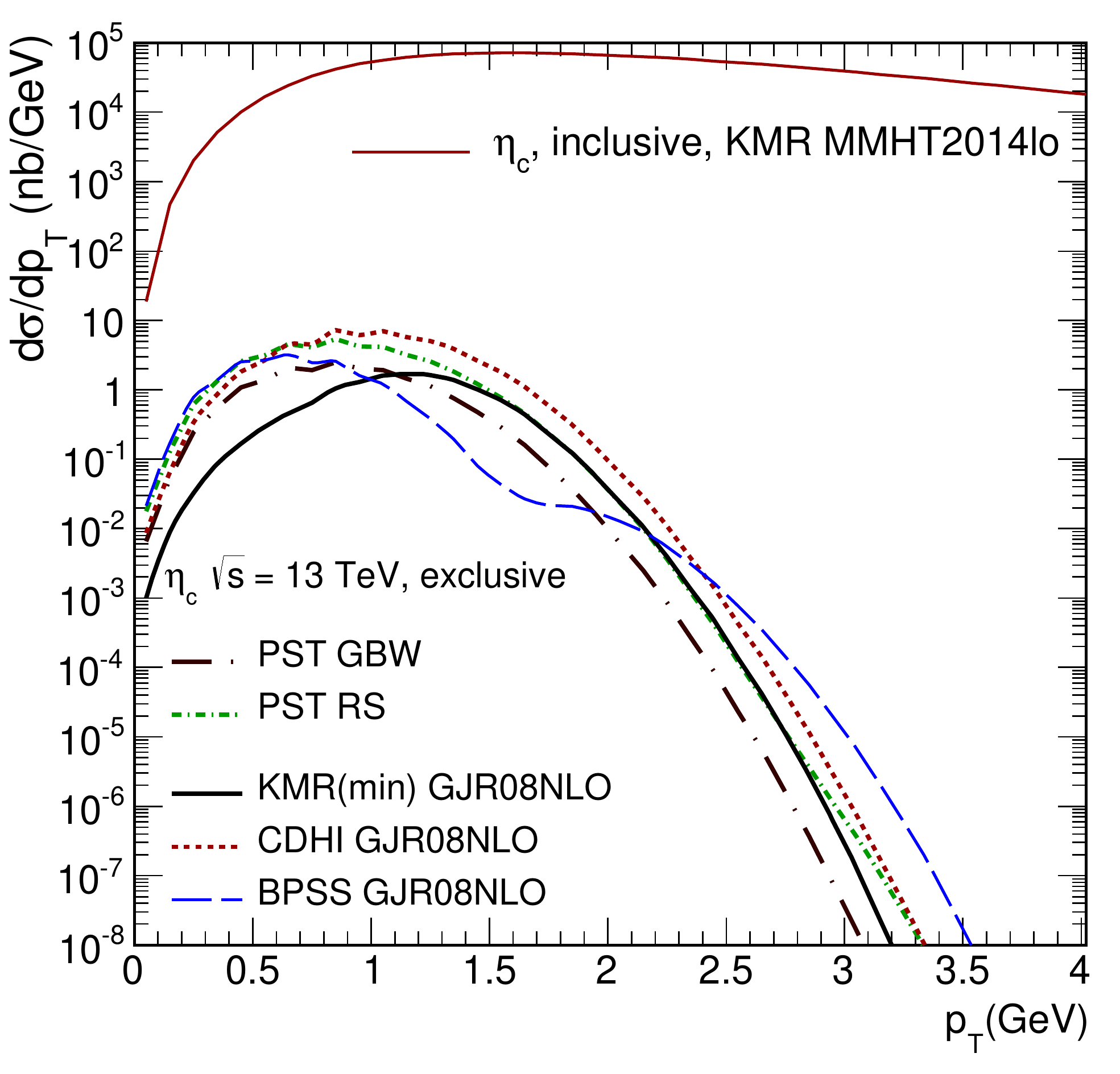}\\
     c)   \includegraphics[width = 0.4\textwidth]{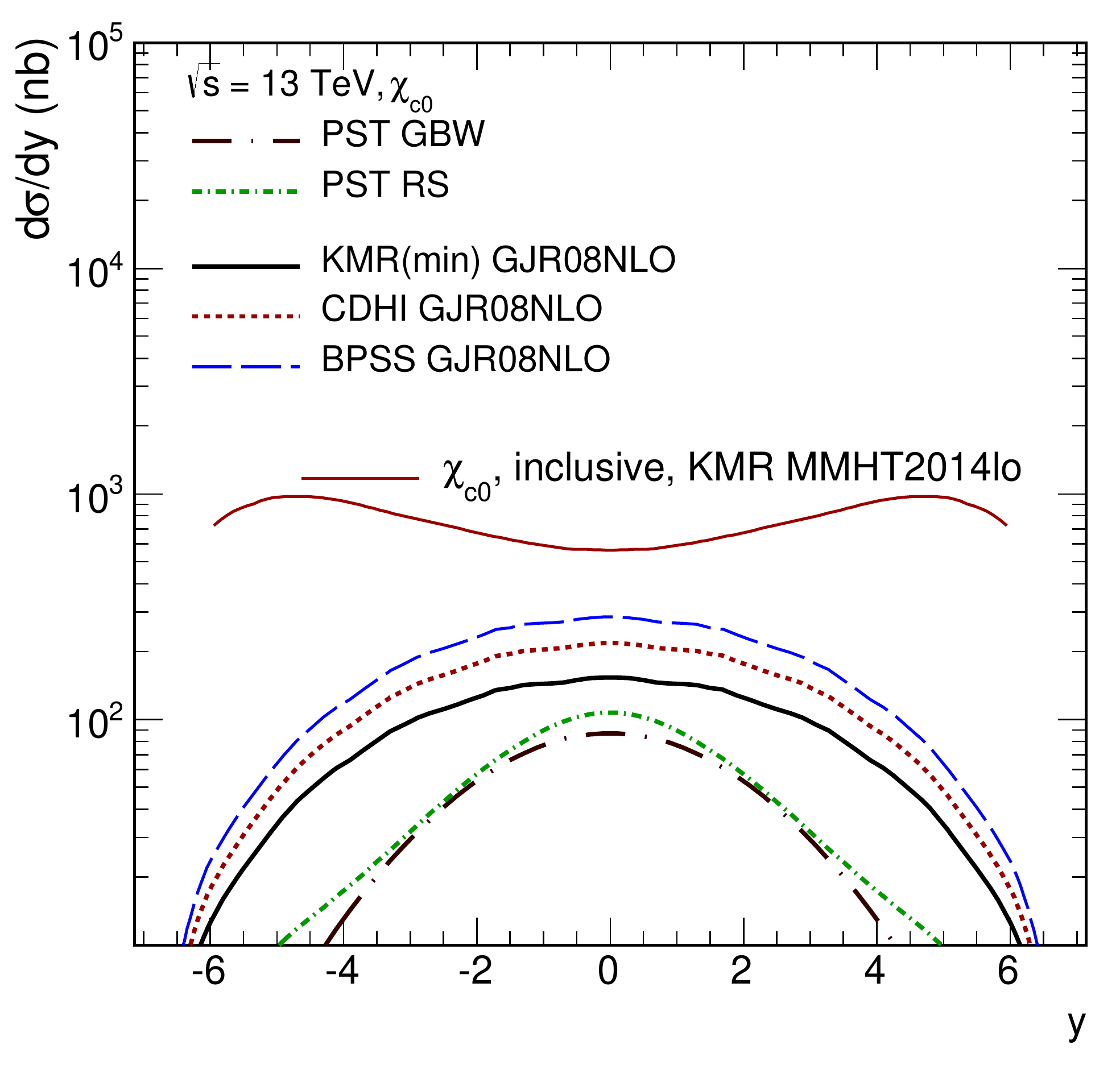}
    d) \includegraphics[width = 0.4\textwidth]{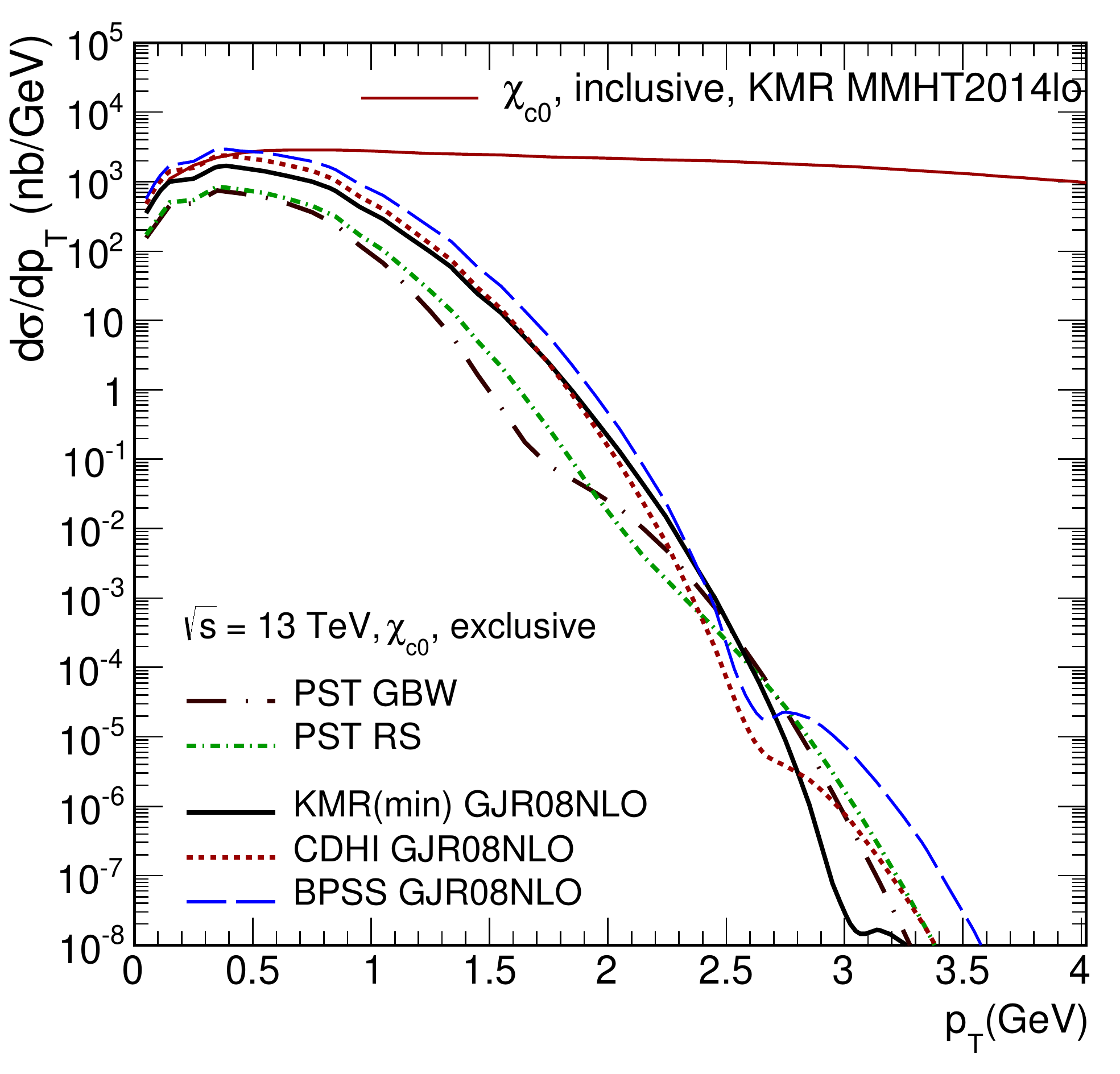}
    \caption{a) Rapidity dependent cross section and b) transverse momentum dependent cross section for $\eta_c$ CEP for various UGD prescriptions; c) \& d): the analogous cross sections for $\chi_{c0}$ CEP. Also shown are the cross sections for the relevant inclusive (nondiffractive) cross section.}
    \label{fig:cross-sec}
\end{figure}

\begin{figure}
    \centering
   a) \includegraphics[width = 0.4\textwidth]{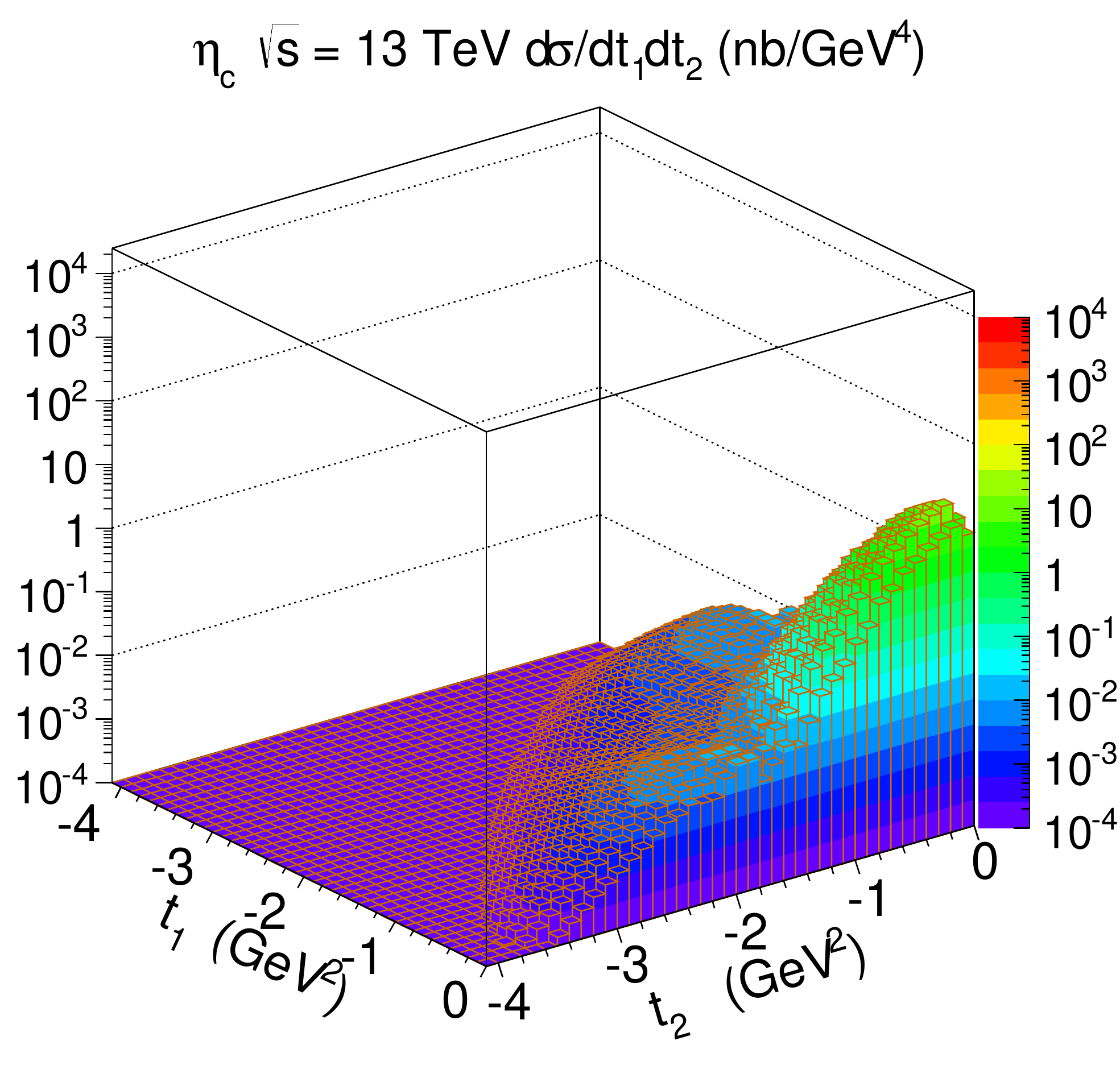}
  b)   \includegraphics[width = 0.4\textwidth]{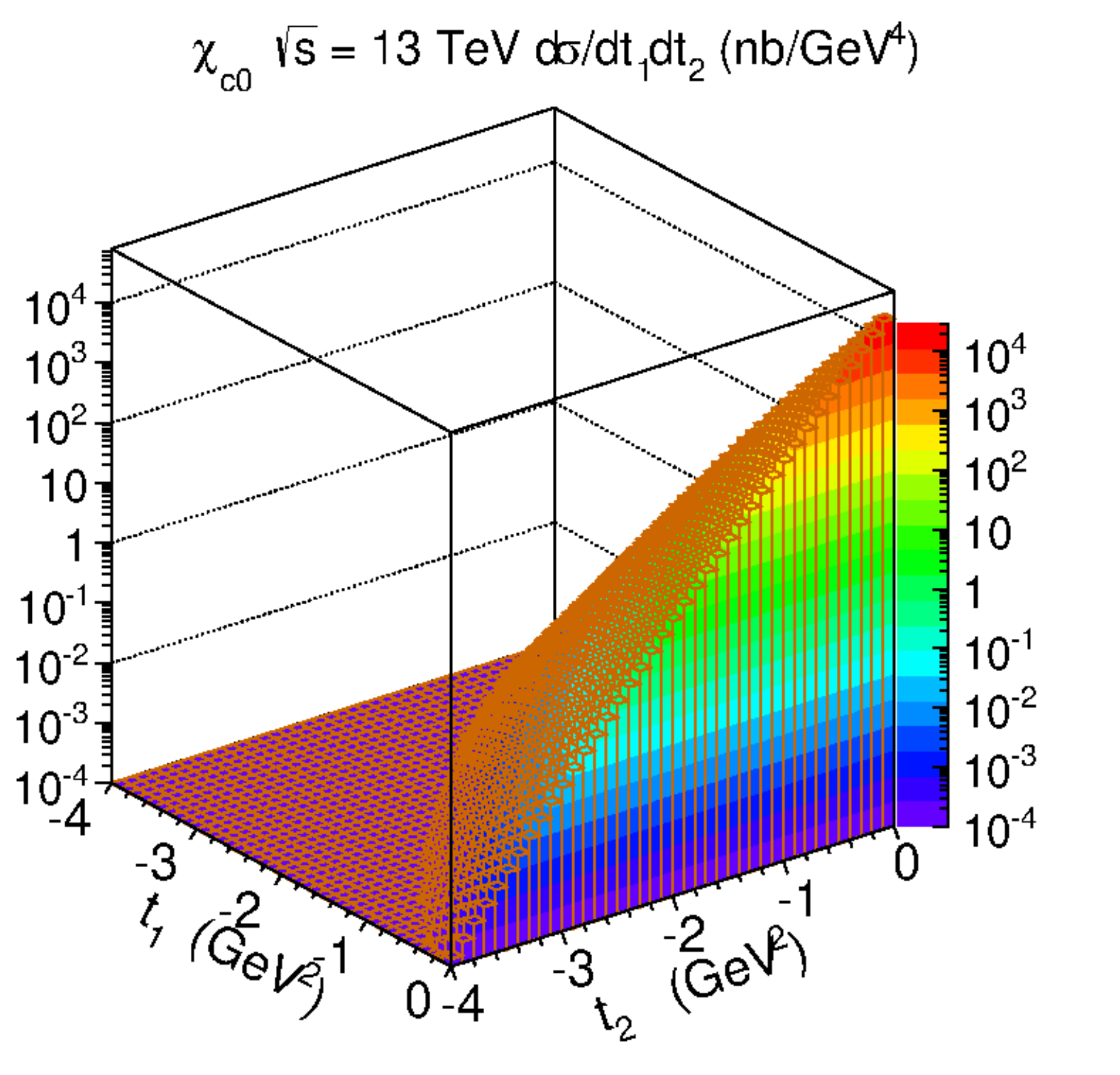}
    \caption{Cross section for a) $\eta_c$ and b) $\chi_{c0}$ CEP differential in Mandelstam-$t_{1},t_{2}$. Here we used the PST prescription.}
    \label{fig:t1t2}
\end{figure}

For the production of bound states of heavy quarks, the quark mass provides a hard scale, and one may attempt a pQCD formulation of the CEP process. The Durham group have proposed a factorization of the CEP amplitude indicated graphically in Fig.~\ref{fig:diagram} -- for a review, see \cite{Harland-Lang:2014lxa}. 
The production of the quarkonium proceeds through the fusion of two gluons. Another gluon -- the so-called screening gluon -- is exchanged between protons and ensures that the $t$--channel exchange is in the color-singlet.
The factorization formula for the CEP amplitude reads:
	\begin{eqnarray}
			{\cal M} = \frac{i s}{4 \pi^2} {\delta_{c_1 c_2} \over N_c^2-1} 
			\int d^2\bQQ \; {\cal V}^{c_1c_2}(\bq_1, \bq_2)
			{{\cal F}_{g}^{\rm off}(x_1, x', \bQQ^2, \bq_{1}^{2}, \mu^2, t_1) 
		   {\cal F}_{g}^{\rm off}(x_2, x', \bQQ^2, \bq_{2}^{2},\mu^2 , t_2) \over
		 \bQQ^2 \bq_{1}^{2} \bq_{2}^{2}} \, ,
			\end{eqnarray}
where ${\cal V}^{c_1c_2}(\bq_1, \bq_2)$ is the vertex that describes the $g^* g^* \to \chi_c$ or $g^* g^* \to \eta_c$ transitions. It is contracted with the polarization vectors of off-shell gluons which are represented by the light-like vectors $n_\mu^{\pm}$:
\begin{eqnarray}
n^+_\mu n^-_\nu {\cal V}^{ab}_{\mu \nu} (\bq_1, \bq_2) = {4 \pi \alpha_s \over \sqrt{N_c}} \, 
\delta^{ab} {\cal T} \, .
\end{eqnarray}  
For the case of the $\chi_{c0}$, the transition amplitude can be decomposed into two form-factors
\begin{eqnarray}
  {\cal T} &=&  |\bq_1| |\bq_2| G_{1}(\bq_1^2,\bq_2^2) + 
  (\bq_1 \cdot \bq_2) G_{2}(\bq_1^2,\bq_2^2) \, ,
\end{eqnarray}
for which we have derived a representation in terms of light-front wave functions (LFWFs) \cite{Babiarz:2020jkh}:
\begin{eqnarray}
 G_1(\bq_1^2, \bq_2^2) &=& |\bq_1||\bq_2| \,  {4m_c \over \bq_2^2}  \int {dz d^2\bk \over z(1-z) 16 \pi^3}  \psi_{\chi_{c0}}(z,\bk)  \,  2z(1-z) (2z-1) \Big[ {1 \over \bl_A^2 + \varepsilon^2} - {1 \over \bl_B^2 + \varepsilon^2}\Big] \nonumber \\
 G_2(\bq_1^2,\bq_2^2) &=&
 4m_c \int {dz d^2\bk \over z(1-z) 16 \pi^3}  \psi_{\chi_{c0}}(z,\bk) \Big[ {1 -z \over \bl_A^2 + \varepsilon^2} + {z \over \bl_B^2 + \varepsilon^2}\Big] \nonumber \\
 &+& {4 m_c \over \bq_2^2} \int {dz d^2\bk \over z(1-z) 16 \pi^3}  \psi_{\chi_{c0}} (z,\bk) 4z(1-z) 
 \Big[ {\bq_2\cdot \bl_A \over \bl_A^2 + \varepsilon^2} - {\bq_2 \cdot \bl_B \over \bl_B^2 + \varepsilon^2}\Big] \, ,
 \label{eq:G1_G2}
\end{eqnarray}
with $\bl_A = \bk - (1-z) \bq_2, \bl_B = \bk + z \bq_2$ and $\varepsilon^2 = m_c^2 + z(1-z) \bq_1^2$. The LFWFs are obtained using the well-known Terent'ev prescription from potential model rest frame wave functions.
For the case of the pseudoscalar $\eta_c$, there is only one form factor,
\begin{eqnarray}
  {\cal T} &=&  (-i)  [\bq_1,\bq_2] I(\bq_1^2,\bq_2^2) \,  ,
\end{eqnarray}
which reads \cite{Babiarz:2019sfa}:
\begin{eqnarray*}
I(\bq_1^2,\bq_2^2) &=&   4 m_c 
\int {dz d^2 \bk \over z(1-z) 16 \pi^3} \psi_{\eta_c}(z,\bk) 
\Big\{ 
{1-z \over \bl_A^2 + \varepsilon^2}
+ {z \over \bl_B^2 + \varepsilon^2}
\Big\} \, .
\end{eqnarray*}

	\begin{eqnarray}
					{\cal F}_{g, {\rm KMR}}^{\rm off}(x, x', \bQQ^2, \bq^2,\mu^2 , t) = 
					R_g \frac{d}{{d {\rm ln}}\bq^2}\Big[xg(x,\bq^2)
					\sqrt{T_g(\bq^2,\mu^2)}\Big]_{\bq^2 = \bQQ^2} \, \cdot F(t)  \,,
					\end{eqnarray}
The coupling of gluons to protons is described by a generalized unintegrated gluon distribution (UGD). Here several prescriptions exist in the literature: 
	\begin{eqnarray}
  {\cal F}^{\rm off}_{g,{\rm CDHI}} (x, x', \bQQ^2, \bq^2,\mu^2 , t) = 
	R_g \Big[{\partial \over \partial  \log \bar Q^2} 
	\sqrt{T_g(\bar Q^2,\mu^2)}
	\, xg(x,\bar Q^2) \Big] \cdot 
					{2 \bQQ^2 \bq^2 \over \bQQ^4 + \bq^4} \cdot F(t)\,, 
					\end{eqnarray}
with $ \bar Q^2 = (\bQQ^2 + \bq^2)/ 2$
	\begin{eqnarray}
		 {\cal F}^{\rm off}_{g,{\rm PST}}(x, x', \bQQ^2, \bq^2,\mu^2 , t_i) = 
		\sqrt{\bQQ^2 f^{\rm GBW}_g(x',\bQQ^2)
    \bq^2 f^{\rm GBW}_g(x,\bq^2)}\, \sqrt{T_g(\bq^2,\mu^2)}\,\cdot F(t) \,, 
    \end{eqnarray}
for more details, see \cite{Babiarz:2020jkh}. The factor $R_g$ takes into account the so-called skewedness correction, which comes from the fact, that the screening gluon carries a much smaller $x$ than the fusing gluons. In Fig.~\ref{fig:cross-sec} we show rapidity and transverse momentum dependent cross sections for $\eta_c$ and $\chi_{c0}$ CEP for different prescriptions and choices of gluon distributions. In Fig.~\ref{fig:t1t2} distributions 
in Mandelstam-$t_1,t_2$ are shown.
We observe the forward dip for $\eta_c$ and peak for $\chi_{c0}$
at small $t_{1,2}$.

\subsection{Absorptive corrections}
The results shown up to now do not contain the gap survival factors, which encode the effect of absorptive corrections. We estimate the latter in an admittedly crude elastic rescattering approximation. The full amplitude is written as
\begin{eqnarray}
{\cal M}(Y,y,\bp_1,\bp_2) = {\cal M}^{(0)}(Y,y,\bp_1, \bp_2) - \delta {\cal M}(Y,y,\bp_1,\bp_2) \, ,
\end{eqnarray}
with the Born amplitude described above, and the absorptive correction being:
\begin{eqnarray}
\delta {\cal M}(Y,0,\bp_1,\bp_2) &=&
\int {d^2 \bk \over 2 (2 \pi)^2}
\, T(s,\bk)  \exp\Big( -\half B_D (\bp_1 + \bk)^2 \Big)
\exp\Big( -\half B_D (\bp_2-\bk)^2 \Big) \nonumber \\
&\times&  V(\bp_1+ \bk,\bp_2-\bk) \, , 
\end{eqnarray}
with an effective vertex $V$ that describes the fusion of two Pomerons into the meson. It is adjusted to the Born results of the previous sections together with parameter $B_D$.
Above,
\begin{eqnarray}
 T(s,\bk) = \sigma^{pp}_{\rm tot}(s) \, \exp\Big(-\half B_{\rm el}(s) \bk^2 \Big) \, ,
\end{eqnarray}
is the elastic $pp$ amplitude.
We show the gap survival factor 
\begin{eqnarray}
S^2 \equiv  \frac{d \sigma/dy \Big|_{y=0}} {d \sigma_{\rm Born}/  dy \Big|_{y=0}} \, .
\end{eqnarray}
in Table~\ref{table:1} for $\chi_{c0}$ and in Table~\ref{table:2} for $\eta_c$.

\begin{table}[]
    \centering
    \begin{tabular}{l|c|c|c}
    \hline
    \hline
    $\chi_{c0}$ & $\frac{d\sigma}{dy}_{\rm tot} |_{y = 0}$ [nb] & $\frac{d\sigma}{dy}_{\rm tot}^{\rm abs}|_{y = 0}$ [nb]& $S^2_{y=0}$\\
    \hline
    PST GBW           & 17 & 3.7 & 0.22\\
    PST RS            & 21 & 4.5 & 0.21\\
    CDHI GJR08NLO     & 42 & 7.5 & 0.18\\
    KMR  GJR08NLO     & 29 & 3.7 & 0.13\\
    BPSS GJR08NLO     & 61 & 8.0 & 0.13\\
    \hline
       \end{tabular}
       \caption{Gap survival factors for $\chi_{c0}$ CEP for various prescriptions 
       and UGD choices.}
\label{table:1}
\end{table}
\begin{table}[]
    \centering
    \begin{tabular}{l|c|c|c}
         \hline
         \hline
     $\eta_c$ &$\frac{d\sigma}{dy}_{\rm tot}|_{y = 0}$ [nb] & $\frac{d\sigma}{dy}_{\rm tot}^{\rm abs}|_{y = 0}$ [nb] & $S^2_{y=0}$\\
     \hline
    PST GBW       &  $1.8\times10^{-2}$ & $3.9\times10^{-3}$ & 0.22\\
    PST RS        &  $9.0\times10^{-3}$ & $1.9\times10^{-3}$ & 0.21\\
    CDHI GJR08NLO &  $1.8\times10^{-1}$ & $4.0\times10^{-2}$ & 0.22 \\
    KMR  GJR08NLO &  $1.3\times10^{-1}$ & $3.0\times10^{-2}$ & 0.23\\
    BPSS GJR08NLO & $5.8\times10^{-2}$ & $2.2\times10^{-2}$ & 0.38\\
    \hline
    \end{tabular}
    \caption{Gap survival factors for $\eta_c$ CEP for various prescriptions and UGD choices.}
    \label{table:2}
\end{table}

\section{Conclusion}
We have revisited the pQCD formulation of CEP in the example of the production of spinless quarkonia \cite{Babiarz:2020jhy}. The case of the pseudoscalar $\eta_c$ was calculated for the first time.
The novelty consists of the treatment of the transition amplitude for $g^* g^* \rightarrow \eta_c$ and $g^* g^* \rightarrow \chi_{c0}$ which was calculated \cite{Babiarz:2019sfa,Babiarz:2020jkh} using the light-cone wave functions of $c\bar{c}$ states in the framework of potential models.
It turns out that the CEP processes in proton-proton collisions studied by us are sensitive to rather low momentum scales. This is the case  especially for the $\eta_{c}$, and  is responsible for the main uncertainties in the results.
 We consequently proposed a way to calculate the soft effects (in
the region of small gluon transverse momenta) using UGDs 
obtained from color dipole models
and a simple (PST)
prescription for its off-diagonal extrapolation.
In our treatment of absorptive corrections, we restricted
ourselves to the so-called elastic rescattering correction.
Depending on the UGD
used, we obtain for the $\chi_c$ the gap survival values of
$S^2 \,  = (0.13 - 0.21)$, while for the $\eta_c$ production, they are
somewhat higher, $S^2 \, =  (0.21 - 0.38)$.

\section*{Acknowledgements}
\paragraph{Funding information}
The  work  reported here was  partially  supported  by  the  Polish  National  Science  Center (NCN) grant  UMO-2018/31/B/ST2/03537, and by
the Center for Innovation and Transfer of Natural Sciences and  Engineering  Knowledge  in  Rzesz\'ow. I.B. was partially supported by  the Polish  National Agency  for  Academic  Exchange (NAWA) under  Contract  No. PPN/IWA/2018/1/00031/U/0001.



\bibliography{SciPost_Example_BiBTeX_File.bib}

\nolinenumbers

\end{document}